\documentclass[twocolumn]{aastex62}
\usepackage{amssymb}
\usepackage{array,multirow}
\usepackage{enumerate}
\usepackage{bm}

\usepackage[flushleft]{threeparttable}

\newcommand{\degree}{\ifmmode {^{\circ}\ }\else$^{\circ}$\fi}

\newcommand{\Msun}{\ifmmode {M_{\odot}}\else${M_{\odot}}$\fi}
\newcommand{\Rsun}{\ifmmode {R_{\odot}}\else${R_{\odot}}$\fi}
\newcommand{\Porb}{\ifmmode {P_{\rm orb}}\else${P_{\rm orb}}$\fi}
\newcommand{\Pspin}{\ifmmode {P_{\rm spin}}\else${P_{\rm spin}}$\fi}
\newcommand{\Pdot}{\ifmmode {\dot{P}_{\rm spin}}\else${\dot{P}_{\rm spin}}$\fi}
\newcommand{\age}{\ifmmode {\tau_{\rm c}}\else${\tau_{\rm c}}$\fi}

\newcommand{\rL}{\rho_\Lambda}

\newcommand{\CC}{\Lambda}

\newcommand{\Omo}{\Omega_{m 0}}

\newcommand{\rco}{\rho_{c 0}}
\newcommand{\rmo}{\rho_{m 0}}

\newcommand{\pL}{p_{\CC}}

\newcommand{\oD}{\omega_{\rm BD}}

\newcommand{\dpsi}{\dot{\psi}}
\newcommand{\ddpsi}{\ddot{\psi}}

\newcommand{\fracdpsipsi}{\frac{\dpsi}{\psi}}

\newcommand{\rBD}{\rho_{\rm BD}}
\newcommand{\pBD}{p_{\rm BD}}
\newcommand{\wBD}{\omega_{\rm BD}}
\newcommand{\eBD}{\epsilon_{\rm BD}}
\newcommand{\rT}{\rho_{\rm T}}
\newcommand{\pT}{p_{\rm T}}
\newcommand{\weff}{w_{\rm eff}}

\newcommand{\newtext}[1]{{\textcolor{black}{#1}}}

\shorttitle{Brans-Dicke gravity with a cosmological constant}
\shortauthors{Sol\`{a} Peracaula, J., G\'omez-Valent, A., de Cruz P\'{e}rez, J. \& Moreno-Pulido, C.}

\begin{document}

\title{\bf Brans-Dicke gravity with a cosmological constant  smoothes out  $\Lambda$CDM tensions}

\author{Joan Sol\`{a} Peracaula}
\affil{Departament de F\'\i sica Qu\`antica i Astrof\'\i sica, and Institute of Cosmos Sciences (ICCUB),
Universitat de Barcelona, \\ Avinguda Diagonal 647, E-08028 Barcelona, Catalonia, Spain}

\author{Adri\`{a} G\'omez-Valent}
\affil{ Institut f\"ur Theoretische Physik, Ruprecht-Karls-Universit\"at Heidelberg,
Philosophenweg 16, 69120 Heidelberg, Germany}
\affil{Departament de F\'\i sica Qu\`antica i Astrof\'\i sica, and Institute of Cosmos Sciences (ICCUB),
Universitat de Barcelona, \\ Avinguda Diagonal 647, E-08028 Barcelona, Catalonia, Spain}

\author{Javier de Cruz P\'{e}rez}
\affil{Department of Physics, Kansas State University,
116 Cardwell Hall, Manhattan, KS 66506, USA}
\affil{Departament de F\'\i sica Qu\`antica i Astrof\'\i sica, and Institute of Cosmos Sciences (ICCUB),
Universitat de Barcelona, \\ Avinguda Diagonal 647, E-08028 Barcelona, Catalonia, Spain}

\author{Cristian Moreno-Pulido}
\affil{Departament de F\'\i sica Qu\`antica i Astrof\'\i sica, and Institute of Cosmos Sciences (ICCUB),
Universitat de Barcelona, \\ Avinguda Diagonal 647, E-08028 Barcelona, Catalonia, Spain}

\begin{abstract}
We analyze Brans-Dicke gravity with a cosmological
constant, $\Lambda$, and cold dark matter (BD-$\Lambda$CDM for short) in the light of the latest cosmological
observations on distant supernovae, Hubble rate measurements at different redshifts, baryonic acoustic oscillations, large scale structure formation data, gravitational weak-lensing and
the cosmic microwave background under full Planck 2015 CMB likelihood.  Our analysis includes
both the background and perturbations equations. We find that
BD-$\Lambda$CDM is observationally favored as compared to
the concordance $\Lambda$CDM model, which is traditionally defined within General Relativity (GR). In particular, some
well-known persisting tensions of the $\Lambda$CDM with the data, such as the excess in the mass fluctuation amplitude $\sigma_8$
and specially the acute $H_0$-tension with the local measurements, essentially disappear in this context. Furthermore, viewed from the GR standpoint,  BD-$\Lambda$CDM cosmology  mimics quintessence at $\gtrsim3\sigma$ c.l. near our time.
\end{abstract}


\section{Introduction}
\label{sec:intro}
Brans \&  Dicke (BD)  theory is the first historical attempt to extend Einstein's General Relativity (GR) by promoting the Newtonian coupling constant $G_N$ into a variable one in the cosmic time, $G(t)$\,\citep{BD}. In addition to the ordinary gravitational field, it  introduces a new (scalar) field, $\psi$, and a new parameter, $\wBD$.  The effective gravitational coupling $G(t)$ varies as the inverse of $\psi(t)$, and to recover the excellent description of the gravitational phenomena by  GR,  one expects that  $\wBD$ must be sufficiently large in magnitude.

Different experiments in the Solar System and cosmological probes have been able to put stringent bounds on $\wBD$\,\citep{Will2006,Li2013,Umilta2015,Ballardini2016}. We cannot exclude, however,  that the BD behavior at the cosmological scales is different from that which applies in our astrophysical neighborhood\,\citep{Avilez2014,Clifton2012} owing to the possible existence of screening mechanisms, resulting in softer bounds.

At the cosmological level, the most successful paradigm based on GR  is the $\CC$CDM model, which is the standard or `concordance' model of  cosmology\,\citep{PeeblesBook1993}. It assumes the existence of dark matter and a cosmological constant, $\CC$, in addition to other characteristic ingredients of the universe, such as baryons,  photons and neutrinos.  A positive $\CC$ is introduced as the canonical explanation for the observed accelerated expansion of the universe\,\citep{SNIaRiess,SNIaPerl}. For lack of a better  physical explanation, the parameter $\CC$ in the concordance model is associated to the energy density of vacuum, as follows:
$\rL=\CC/(8\pi\,G_N)$.
Herein  we would like  to use the increasingly precise observations on distant supernovae (SnIa), Hubble rate data $H(z)$, baryonic acoustic oscillations (BAO), large scale structure formation (LSS): redshift-space distortions (RSDs) and gravitational weak-lensing (WL); and the cosmic microwave background (CMB), with the aim of testing if we can be sensitive to phenomenological  differences between GR and BD.  Specifically, we aim at checking if some of the well-known discrepancies or tensions currently afflicting the (GR-based)  $\CC$CDM can be smoothed out in the context of the BD-based one, or  BD-$\CC$CDM.

\section{BD-$\CC$CDM cosmology}
\label{sec:motivation}

The action and field equations for BD-gravity are well-known\,\citep{BD}. They involve the ordinary (tensor) gravitational field of GR, $g_{\mu\nu}$, but  also the scalar BD-field $\psi$ (of dimension two in natural units, $\hbar=c=1$) and the (dimensionless) BD-parameter, $\oD$.   As in the original formulation, we assume  no potential for $\psi$, but we  include the cosmological constant, $\CC$, as a fundamental ingredient of the theory. In fact, we wish to consider the same matter and vacuum components as in the concordance $\CC$CDM, except that we replace the GR paradigm by the BD one. The effective gravitational coupling in the latter,  $G\equiv1/\psi$,  is no longer constant but varies slowly  as $\psi$ itself.    The vacuum energy density and pressure are $\rL=\CC/\kappa^2$ and $\pL=-\rL$, respectively, where for convenience we have introduced  $\kappa^2\equiv  8\pi{G_N}$ and  $G_N\equiv 1/M^2_P$, with $M_P =1.22\times 10^{19}$GeV  the Planck mass in natural units.  Adding them to the corresponding  matter density  $\rho=\sum_i\rho_i$ and pressure $p=\sum_ip_i$ (which may involve both relativistic and nonrelativistic components) we can form the total density and pressure of the combined system of matter and vacuum: $\rT=\rho+\rL$ and  $\pT=p-\rL$.  In the following we focus on the flat FLRW metric only,  i.e. $ds^2=-dt^2 + a^2\delta_{ij}dx^idx^j$, where $a(t)$ is the scale factor as a function of the cosmic time, and we define the usual Hubble rate $H=\dot{a}/a$  (dot denoting $d/dt$).

%
%
\begin{table*}[t!]
\centering
\begin{tabular}
{|c  ||c | c ||  c | c |   }
 \multicolumn{1}{c}{} & \multicolumn{2}{c}{DS1} & \multicolumn{2}{c}{DS2}
\\\hline
{\small Parameter} & {\small $\Lambda$CDM}  & {\small BD-$\CC$CDM} & {\small $\Lambda$CDM}  &  {\small BD-$\CC$CDM}
\\\hline
{\small $H_0$ (km/s/Mpc)}  & {\small $68.65^{+0.38}_{-0.40}$} & {\small $71.03^{+0.91}_{-0.86}$} & {\small $68.69^{+0.38}_{-0.39}$}  & {\small $72.00^{+1.00}_{-1.10}$}
\\\hline
{\small $\Omo$} & {\small $0.2955\pm 0.0048$}  & {\small  $0.2742\pm 0.0077$} & {\small $0.2950\pm 0.0047$}  & {\small  $0.2665\pm 0.0084$}
\\\hline
{\small $\Omega_{b0}$} & {\small $0.0476\pm 0.0004$}  & {\small $0.0453\pm 0.0010$} & {\small $0.0476\pm 0.0004$}  &  {\small $0.0443\pm 0.0012$}
\\\hline
{\small $\tau$} & {{\small$0.063^{+0.010}_{-0.012}$}} & {{\small$0.081^{+0.015}_{-0.018}$}} & {{\small$0.063^{+0.010}_{-0.011}$}}  &   {{\small$0.084\pm 0.018$}}
\\\hline
$n_s$ & {{\small$0.9700^{+0.0038}_{-0.0040}$}}  & {{\small$0.9891^{+0.0070}_{-0.0082}$}} & {{\small$0.9704\pm 0.0038$}} &   {{\small$0.9945^{+0.0081}_{-0.0086}$}}
\\\hline
{\small$\sigma_8(0)$}  & {{\small$0.804^{+0.007}_{-0.009}$}}  & {{\small$0.801\pm 0.010$}} & {{\small$0.804^{+0.007}_{-0.008}$}}  &   {{\small$0.803^{+0.011}_{-0.010}$}}
\\\hline
$\eBD$ & - & {{\small $-0.00277^{+0.00170}_{-0.00154}$}} & - &   {{\small $-0.00315^{+0.00168}_{-0.00175}$}}
\\\hline
$\varphi_{ini}$ & - & {{\small $0.924^{+0.021}_{-0.023}$}} & - &    {{\small $0.901^{+0.026}_{-0.025}$}}
\\\hline
$\varphi(0)$ & - & {{\small $0.904^{+0.028}_{-0.029}$}} & - &    {{\small $0.879\pm 0.032$}}
\\\hline
$w_{eff}(0)$ & -1 & {{\small $-0.961^{+0.012}_{-0.011}$}} & -1 &    {{\small $-0.951^{+0.012}_{-0.013}$}}
\\\hline
$\dot{G}(0)/G(0) (10^{-13}yr^{-1})$ & - & {{\small $3.149^{+1.741}_{-1.924}$}} & - &    {{\small $3.625^{+1.994}_{-1.954}$}}
\\\hline
$\Delta$DIC ($\Delta$AIC) & - & {{\small $8.34$ ($7.72$)}} & - &    {{\small $9.89$ ($9.94$)}}
\\\hline
\end{tabular}
\label{tableFit1}
\caption{\scriptsize  The mean fit values and $68.3\%$ confidence limits for  the considered models ($\CC$CDM and BD-$\CC$CDM) using two datasets: 1) DS1, i.e. SnIa+$H(z)$+BAO+LSS+CMB+R19 with full Planck 2015 CMB likelihood (first block); and 2) DS2, based on the subset BAO+LSS+CMB+R19 (second block).   In all cases a massive neutrino of  $0.06$ eV has been included. First, we display the fitting values for the conventional free parameters: Hubble parameter, $H_0$, the total nonrelativistic matter parameter $\Omo$, and the baryonic part $\Omega_{b0}$, the reionization optical depth $\tau$, the spectral index $n_s$ of the primordial power spectrum, and, for convenience, instead of the amplitude $A_s$ of the spectrum we list the value of $\sigma_8(0)$.  The specific free parameters of the BD model are $\eBD$ and $\varphi_{ini}$, see text. \newtext{Owing to their significance for our analysis, we quote their values with the error bars at $1\sigma$, $2\sigma$ and $3\sigma$ c.l., to wit:  $\eBD = -0.00277^{+0.00170+0.00312+0.00458}_{-0.00154-0.00324-0.00484}$ and $\varphi_{ini} = 0.924^{+0.021+0.044+0.066}_{-0.023-0.042-0.064}$ for the DS1 scenario, and $\eBD = -0.00315^{+0.00168+0.00338+0.00515}_{-0.00175-0.00341-0.00488}$ and $\varphi_{ini} = 0.901^{+0.026+0.052+0.077}_{-0.025-0.050-0.080}$ for the DS2 one.}
Finally, we include the computed values of three parameters at present: the value of the BD-field,  $\varphi(0)$, the effective EoS  $w_{eff}(0)$, Eq.\, (\ref{BDEoS}), and the relative time variation of $G$, $\dot{G}(0)/G(0)$.  In the last row we compare the fit qualities by displaying the differences of the DIC and AIC information criteria between  $\Lambda$CDM and its BD counterpart for the two data sets under consideration. BD appears to be preferred by the data according to these criteria. See the text for further details.}
\end{table*}

\subsection{Field equations}
\label{sec:BDinterpretation}
 With the above notation, we can write down the BD-field equations in the presence of matter and vacuum components as follows:
\begin{equation}
3H^2 + 3H\fracdpsipsi -\frac{\wBD}{2}\left(\fracdpsipsi\right)^2 = \frac{8\pi}{\psi}\rT\,,\label{BDFriedmannEquation}
\end{equation}
\begin{equation}
2\dot{H} + 3H^2 + \frac{\ddpsi}{\psi} + 2H\frac{\dpsi}{\psi} + \frac{\wBD}{2}\left(\frac{\dpsi}{\psi}\right)^2 = -\frac{8\pi}{\psi}\pT\label{BDpressureEquation}
\end{equation}
and
\begin{equation}\label{BDwaveEquation}
\ddpsi +3H\dpsi = \frac{8\pi}{2\wBD +3}(\rT-3\pT)\,.
\end{equation}
For constant $\psi$, the first two equations reduce to the Friedmann and pressure equations of GR,  respectively, with $G=1/\psi=G_N$, and the third requires $|\oD|\to\infty$ for consistency. By combining the above equations one  finds a local covariant conservation law which is identical to that of  GR since there is no interaction between matter and the BD-field, namely
$\dot\rho_T + 3H(\rT+\pT)=0$.  Being $\rL=$const. and owing to the equation of state (EoS) of vacuum ($\pL=-\rL$), the previous equation reduces to $\dot{\rho} + 3H(\rho+p)=0$. Taking into account that we  assume separate conservation of the different components, such a law can be conveniently split into a conservation law for each component (baryons, dark matter, neutrinos and photons).   Hearafter we use the following definitions
\begin{equation}\label{definitions}
\varphi(t) \equiv G_N\psi(t)= G_N/G(t)\,,\qquad \eBD \equiv \frac{1}{\wBD},
\end{equation}
where $\varphi$ is the dimensionless BD-field. In this notation,  $G(t)\equiv G(\varphi(t))=G_N/\varphi$ plays the role of effective gravitational coupling in the BD  context at the level of the action and field equations. We should
 {\emph not} expect $G(t)$ to be equal to $G_N$ at the present time ($t_0$); thus, $\varphi(t_0)$ is in general close, but not exactly equal, to $1$. We can recover GR in the limit $\eBD \rightarrow 0$, \newtext{and the $\Lambda$CDM model when $\eBD\rightarrow 0$ and $\varphi\rightarrow 1$.}

\begin{figure*}
\begin{center}
	\includegraphics[width=1.2\columnwidth]{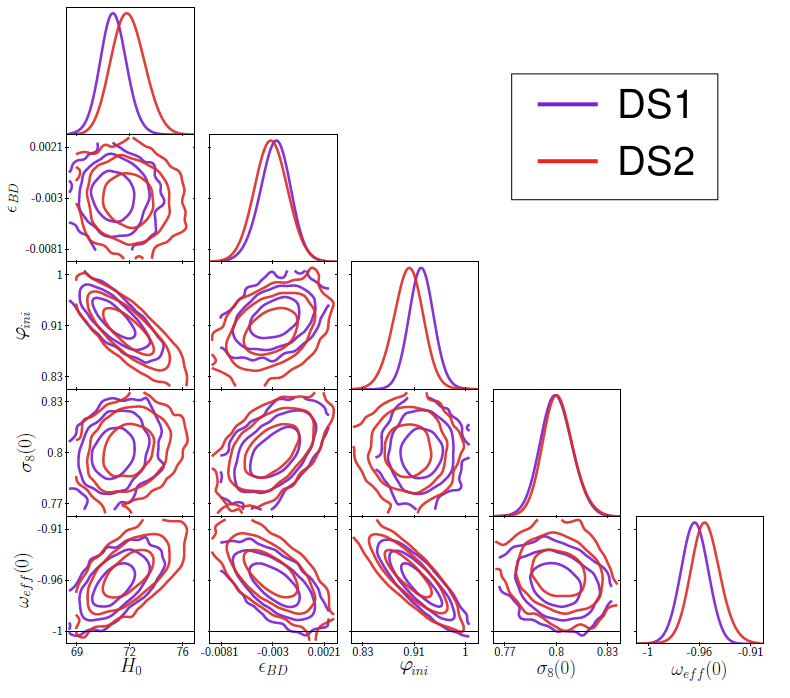}
     \caption{Triangular matrix containing some relevant combinations of two-dimensional marginalized distributions for fitting parameters of the BD model (at $1\sigma$, $2\sigma$ and $3\sigma$ c.l.), together with the corresponding one-dimensional marginalized likelihoods for each parameter. $H_0$ is expressed in km/s/Mpc. We present the contours for the data sets DS1 (in purple) and DS2 (in red) (cf. Table 1 for the numerical fitting results).}
    \label{fig:CLnewV2}
   \end{center}
\end{figure*}

\subsection{Effective equation of state  in the GR-frame}
\label{sec:EffEoS}

It proves revealing to analyze Brans-Dicke cosmology from the standpoint of what we may call the `GR-frame'. The latter is obtained by parameterizing the BD-field equations  as deviations with respect to the usual Friedmann's equations.  For instance, in \cite{GRF2018} and \,\cite{MPLA2018}  it was shown, using an approximate treatment, that such kind of approach leads to mimic `running vacuum', starting from a  cosmological constant. Vacuum dynamics, and in general dynamical dark energy (DE), can be phenomenologically favorable, even if not firmly established yet, see e.g.\,\cite{ApJs,Zhao2017,MNRAS-EPL,ParkRatra2018-19,PDU2019,Martinelli2019} and references therein. It can also be a cure for some of the tensions in the $\CC$CDM\,\citep{Valentino2017,PLB2017}.  Here we undertake a systematic approach and an exact numerical treatment of the BD-field equations in the GR-frame, which will lead to effective dynamical DE. The first step is to rewrite Eq.\,(\ref{BDFriedmannEquation}) as if it were the usual Friedmann equation,
$
3H^2 =\kappa^2(\rT+ \rBD)\,.
$
It is not difficult to show that the effective energy density associated to the BD-field,  $\rBD$, reads
\begin{equation}\label{rhoBD}
\kappa^2\rBD=3H^2\,\Delta\varphi - 3H\dot{\varphi} + \frac{1}{2\eBD}\frac{\dot{\varphi}^2}{\varphi}\,,
\end{equation}
where  $\Delta\varphi\equiv1-\varphi$. Recall our definitions (\ref{definitions}); we have just included all the energy density terms beyond the $\Lambda$CDM  in the expression for $\rBD$.  Such an energy density can therefore  be interpreted as a new  DE component within the GR-frame, which must be added to the vacuum  part $\rL$. The second step is to recast Eq.\,(\ref{BDpressureEquation}) as the usual  Friedmann's  pressure equation:
$
2\dot{H} + 3H^2 = -\kappa^2(\pT+ \pBD)\,.
$
This leads to the following expression for the effective pressure associated to the BD-field (playing the role of `extra DE pressure', in addition to $\pL=-\rL$):
\begin{equation} \label{pBD}
\kappa^2\pBD=- 3H^2\,\Delta\varphi -2\dot{H}\,\Delta\varphi + \ddot{\varphi}+ 2H\dot{\varphi} + \frac{1}{2\eBD}\frac{\dot{\varphi}^2}{\varphi}\,.
\end{equation}
%
Direct calculation from the three field equations (\ref{BDFriedmannEquation})-(\ref{BDwaveEquation}) shows that such a fluid obeys the  additional conservation law
\begin{equation}
\dot{\rho}_{\rm BD}+3H(\rBD+\pBD)= 0.
\end{equation}

One can also cross check it from $\nabla^\mu \tilde{T}^{\rm BD}_{\mu\nu}=0$, where the energy-momentum tensor
 $\tilde{T}^{\rm BD}_{\mu\nu}\equiv-\frac{2}{\sqrt{-g}}\frac{\delta \tilde{S}_{\rm BD}}{\delta g^{\mu \nu}}$ is computed from the part of the full BD-action $S_{\rm BD}$ that remains after we subtract from it the usual Einstein-Hilbert action $S_{\rm EH}$ (with cosmological term) and  the matter action, $S_m$,  i.e.  $\tilde{S}_{\rm BD}=S_{\rm BD}-S_{\rm EH}-S_m$.

From the previous considerations, the BD-fluid can be described by the quantities $\pBD$ and $\rBD$.
However, a more useful picture can be obtained by considering the EoS of the full  `effective DE' in this context. Obviously, it must receive contributions from the BD-fluid and the vacuum. Using the field equations in the GR-frame as discussed above, we find  $\weff= (\pL + \pBD)/(\rL + \rBD)$.  It can be conveniently cast as follows:
\begin{equation}\label{BDEoS}
\weff(t)=-1+\frac{-2\dot{H}\Delta\varphi+f_1(\varphi,\dot{\varphi},\ddot{\varphi})}{\CC+ 3 H^2\,\Delta\varphi+f_2(\varphi,\dot{\varphi})}\,.
\end{equation}
The two functions $f_{1,2}$ need not be specified here, they can be easily computed from the previous formulae. It suffices to say that they are numerically negligible, in absolute value,  as compared to $\dot{H}\Delta\varphi$ and  ${H^2}\Delta\varphi$ since they depend on time derivatives of the slowly varying function $\varphi$. The effective EoS (\ref{BDEoS}) is, obviously, a time-evolving quantity which mimics dynamical DE  in the above effective  GR picture. Since $|\Delta\varphi|$ is small, as confirmed by our analysis (cf. Sec. \ref{sec:Numerical}), let us remark the following interesting situation near our time (i.e. for cosmological redshift  $z\ll1$):
\begin{equation}\label{BDEoSz0}
\weff(z)\simeq-1-\frac{2\dot{H}\Delta\varphi}{\CC}\simeq -1+\Delta\varphi\,\frac{\Omo}{1-\Omo}\,(1+z)^3\,,
\end{equation}
where we have expanded  linearly in $\Delta\varphi$ and expressed the result in terms of the  current value of the matter cosmological parameter $\Omo=\rmo/\rco$. Here $\rco=3H_0^2/\kappa^2$ is the critical density at present and $H_0$ is the Hubble parameter. Clearly, for $\Delta\varphi>0$ (resp. $<0$) we meet quintessence-like (resp. phantom-like) behavior. We shall further discuss this important issue in Sec. \ref{sec:Numerical}.

\section{Structure formation and perturbations}
\label{sec:confrontation with data}
For a full fledged confrontation of the BD-$\CC$CDM model with the observations we need to consider not only its background features but also the implications on the large scale structure (LSS) formation data, which we include in our global fit (cf. Section \ref{sec:Confrontation}). Thereby we need to account for  the matter density perturbations in the BD framework.  We have performed such a nontrivial calculation both within the synchronous and conformal Newtonian gauges.  They render  the same result at deep subhorizon scales ($k^2\gg (aH)^2$), as they indeed should.  Details on such, more technical, part of the analysis will be provided elsewhere in the context of a more complete presentation.   \newtext{Here we just quote the obtained differential equation satisfied by the linear matter density contrast at the mentioned scales, which is fully in accordance with  the analysis of \cite{Boisseau2000}}:
\begin{equation}\label{eq:perturbEq}
\delta_m^{\prime\prime}+\mathcal{H}\delta_m^\prime-\frac{4\pi G_N a^2}{\bar{\varphi}}\bar{\rho}_m\delta_m\left(\frac{2+4\,\eBD}{2+3\,\eBD}\right)=0\,,
\end{equation}
where $\bar{\varphi}$ and $\bar{\rho}_m$ are the mean values of $\varphi$
\begin{figure*}
\begin{center}
	\includegraphics[width=0.5\textwidth]{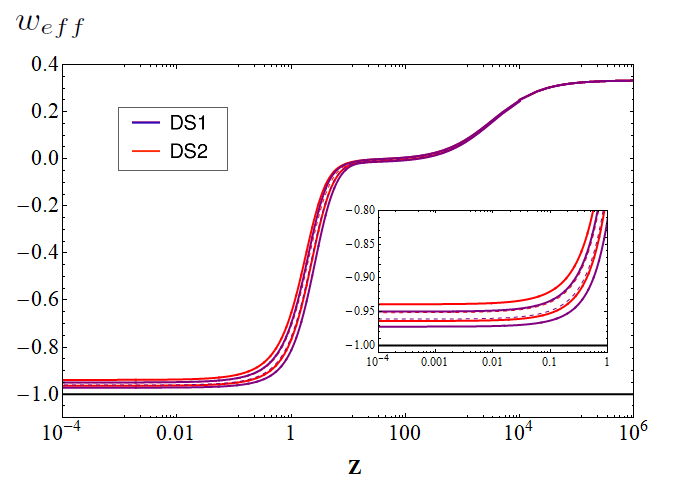}
     \caption{The effective EoS parameter for the BD model and its corresponding $1\sigma$ bands as a function of the redshift,  Eq.\,(\ref{BDEoS}). We plot the results derived under both, dataset DS1 (in purple) and DS2 (in red), in the redshift range $z<10^6$. In black we plot the EoS parameter of the vacuum energy density, i.e. $w_\Lambda =-1$ . During the matter and radiation-dominated eras the EoS of BD-$\CC$CDM tracks the dominant component of the universe. This is no longer true at low redshift near our time. We zoom in the region $z<1$ to better appreciate the existing deviation of the EoS parameter from $-1$. \newtext{With the error bars at $1\sigma$, $2\sigma$ and $3\sigma$ the results are the following: $w_{eff}(0) = -0.961^{+0.012+0.023+0.033}_{-0.011-0.023-0.034}$ for the DS1 scenario, and $w_{eff}(0) = -0.951^{+0.012+0.026+0.041}_{-0.013-0.026-0.039}$ for the DS2 one. These results point to an effective quintessence signal at $\gtrsim3\sigma$ c.l. in both cases.}}
    \label{fig:EoS_BDnew}
    \end{center}
\end{figure*}
%
and of the matter density ${\rho}_m$, respectively, at the cosmological scales in which the linear theory remains valid. Prime denotes differentiation with respect to conformal time  $\eta$ (recall that $dt=a\,d\eta$) and $\mathcal{H}=a'/a$.  It is easy to recognize the corresponding perturbations equation for the $\Lambda$CDM in the limit $\eBD\to 0$ (i.e. $\wBD\to\infty$), as  could be expected. It follows from (\ref{eq:perturbEq}) that the effective value of the gravitational constant driving  the formation of linear structures in the BD theory, at subhorizon scales, is
%
$\frac{G_N}{\bar{\varphi}}\left(\frac{2+4\,\eBD}{2+3\,\eBD}\right)$.
%
Even though a similar relation holds between the local gravitational field created by a spherical mass and the BD-field in the weak-field limit\,\citep{BD},
as indicated in the introduction we take the wider perspective that the BD theory, when applied to the cosmological level, is not restricted by the bounds obtained in the astrophysical neighborhood.  This is one of the traditional scenarios explored in the literature,  the alternative one being the identification of the two domains. Both views are possible \,\citep{Avilez2014} and here we address the less restrictive one, on account of the screening effects that may be produced by the presence of matter in the local domain, as it has been exemplified in other contexts, see e.g.\,\cite{DEBook,Clifton2012} and references therein.

\section{Confrontation with data}
\label{sec:Confrontation}

\subsection{Data}
\label{sec:data}

We use two datasets, which we believe  are helpful to better focus on the origin of the main effects. The first one is labeled  DS1 and contains the cosmological data SnIa+$H(z)$+BAO+LSS+CMB+R19; the second one,  DS2,  is just the subset BAO+LSS+CMB+R19 of the first. The CMB part involves the full Planck 2015 TT+lowP+lensing likelihood\,\citep{Planck2015}. These two datasets are essentially the same as those described in detail in \cite{PDU2019}. Here, however, apart from the cosmic chronometer data on the Hubble rate  $H(z)$, we have included the latest local value of the Hubble parameter,  $H_0 = 74.03 \pm 1.42 $ km s$^{-1}$ Mpc$^{-1}$\, \citep{R19}, which we have denoted R19, based on distance ladder measurements. This is an important additional input, given the significant existing tension ($\sim 4.4\sigma$) of such a value with the Planck results, as discussed in that reference.  In addition, we have added up the independent SnIa data from the DES survey \citep{DESSNIa} to the Pantheon+MCT dataset\,\citep{Scolnic2018,Riess2018} already used in the mentioned reference.  In the BAO sector, we include a new data point from the DES survey \citep{Camacho2018}  and the combined Ly$\alpha$-forest data by\,\citep{Blomqvist2019}. See Tables 4 and 5 of  \cite{PDU2019} for the remaining list of BAO and RSD data.  Note that, in the last two sets, we  include the matter bispectrum data from \cite{GilMarin2016}  (labeled BSP in the aforementioned tables). Finally, we use the WL data from \cite{Hildebrandt2018}.

\subsection{Numerical analysis and model selection}
\label{sec:Numerical}

To compare the theoretical predictions of the models under study with the available observational data, we have made use of the Einstein-Boltzmann code CLASS \citep{CLASS} in combination with the Monte Carlo Markov chain (MCMC) sampler MontePython\,\citep{MontePython}. Apart from the common set of basic $6$ parameters of the concordance model, the BD-$\CC$CDM model involves two more, all of them listed in Table 1.    One of the characteristic BD parameters is obviously  $\eBD$.   However,  in order to solve numerically the system of differential equations (\ref{BDFriedmannEquation})-(\ref{BDwaveEquation}) we use a second fitting parameter, namely the  initial value of the BD-field,  $\varphi_{ini}$,  which is set at   $z_{ini}=10^{14}$.  The evolution of  $\varphi$ proves to be very mild and we naturally take its derivative at that point to be zero. \newtext{We have checked that the small variation induced in the expansion rate at BBN is within bounds\,\citep{Uzan2011}}.  The main fitting results of our analysis are displayed in  Table 1 and Fig.\,1.  In the table, we compare the standard $\CC$CDM model with its  BD-$\CC$CDM counterpart.  Remarkably enough, the contour lines in Fig.\, 1 show a preference for relatively high values of $H_0$, not far from R19,  while keeping $\sigma_8\equiv\sigma_8(0)$ at an intermediate level between Planck measurements\,\citep{Planck2015} and cosmic shear data\,\citep{Hildebrandt2018}, thus smoothing out this tension as well. In Fig.\,1, we provide the matrix containing some relevant combinations of two-dimensional marginalized distributions for fitting parameters of the BD model, together with the corresponding one-dimensional marginalized likelihoods for each parameter. We confirm indeed the softening of the two tensions in view of the fitted values of $H_0$ around  $71-72$ km s$^{-1}$ Mpc$^{-1}$,  which  coexist peacefully with a sufficiently low  $\sigma_8\simeq 0.80$. The residual $H_0$-tension with the local measurement\,\citep{R19} comes down to  $1.8\sigma$ (DS1) and  $1.1\sigma$ (DS2) only.

How about the global quality of our fits? Occam's razor principle can be implemented rigorously with various Bayesian model selection tools, see e.g. \cite{Burnham2002}. In this Letter, we opt for making use of the Deviance \citep{DIC} and Akaike \citep{Akaike} information criteria,  referred to as DIC and AIC, respectively. When a large amount of data is employed,  AIC is simply given by ${\rm AIC}=\chi^2(\hat{\theta})+2n$, where  $n$ is the number of independent fitting parameters and $\hat{\theta}$ is  the collection of their mean values. DIC is a more sophisticated generalization of AIC, being defined as
\begin{equation}
{\rm DIC}=\chi^2(\hat{\theta})+2p_D\,.
\end{equation}
Here $p_D=\overline{\chi^2}-\chi^2(\hat{\theta})$ is the effective number of parameters of the model  and $\overline{\chi^2}$ the mean of the overall $\chi^2$ distribution. DIC is particularly suitable for us, since we can easily compute all the quantities involved directly from the Markov chains and other output generated with MontePython.
Both, DIC and AIC, are reliable provided the posterior distributions are sufficiently Gaussian. This is actually the case here, as reflected in the elliptic shapes of the two-dimensional contours in Fig.\,1 and also in the normal-like appearance of the one-dimensional distributions shown there.
To compare the ability of the $\Lambda$CDM and BD-$\Lambda$CDM models to fit the data, one has to compute the respective differences of DIC and AIC values between the first and second models. They are denoted  $\Delta$DIC  and $\Delta$AIC  in Table 1,
%
%
where we provide the results for both datasets DS1 and DS2.  Since these differences are positive and both  lie in the interval $5<\Delta{\rm AIC},\Delta{\rm DIC}<10$  we conclude, following \cite{Burnham2002,DIC}, that they  point to  strong evidence in favor of BD-$\Lambda$CDM as compared to $\Lambda$CDM.

The following observations are also in order. The fitted  $\eBD$  in Table 1 entails a  large enough value of $|\wBD|={\cal  O}(300) $ such as  to guarantee that the BD-$\CC$CDM model remains sufficiently close to the concordance one, but not so large as to make the BD approach phenomenologically irrelevant.  At the same time, negative values of $\eBD$ are preferred in our numerical analysis. Both features are consistent with previous estimates from  analytical power-law solutions found in \cite{GRF2018} and \,\cite{MPLA2018}. Furthermore, the relative time variation of $G$ at present is found to be positive, $\dot{G}(0)/G(0)\simeq +3\times 10^{-13}\,yr^{-1}$(at roughly $2\sigma$), which indicates `asymptotically free' behavior (i.e. $G$ mildly increasing with the expansion), also consistent with the aforementioned power law-solution.

Last but not least, worth noticing is the behavior of the effective EoS (\ref{BDEoS}), which we plot numerically in Fig.\,2. Being $\Delta\varphi>0$ throughout our analysis (cf. Table 1) we can confirm the expected quintessence-like signal, which we had anticipated with the approximate formula (\ref{BDEoSz0}) near our time ($z\ll1$) -- see the inner plot in that Figure.  As can be seen, a rather conspicuous signal in between $3\sigma$ to $4\sigma$ is obtained, depending on the dataset.  In the opposite end ($z\gg1$), i.e.  deep into the matter- or radiation-dominated epochs, the behavior of (\ref{BDEoS}) is of the form $\weff \simeq -1 - \frac{2}{3}\frac{\dot{H}}{H^2}$ (as  $\CC$ and $f_{1,2}$  become negligible against  $|\dot{H}|\Delta\varphi$ and  ${H^2}\Delta\varphi$).  The characteristic EoS of these epochs is then met in sequence ($\weff\simeq 0, 1/3$).

\section{Conclusions}
\label{sec:conclusions}

BD-cosmology is  based on a  gravity paradigm different from GR, in which the gravitational coupling is mildly evolving with the expansion. This is still compatible with the weak form of the equivalence principle.  However, a new degree of freedom comes on stage. In this Letter, we have used a large body of modern cosmological data to explore the possible impact that such a change of paradigm can have on the overall description of observations, while still keeping the  usual matter and vacuum concepts of the concordance $\CC$CDM model. This can be timely, if we bear in mind the current weaknesses or tensions of the $\CC$CDM  with some observational data, as widely recognized in the literature\,\citep{R19}.  We have found that the BD-$\CC$CDM cosmology can appear  $\CC$CDM-like with, however, a mild time-evolving DE component whose effective EoS mimics quintessence at more than $3\sigma$ c.l. around our time.  The latter acts as a ``smoking gun'' of the underlying BD-dynamics.  Using standard information criteria tools we have confirmed that the statistical quality of the BD fit is strongly preferred to that of a rigid $\CC$-term. Finally, the mass fluctuation amplitude $\sigma_8$ stays at a low enough level and the sharp $H_0$-tension with the local measurements is rendered essentially harmless in this context.\\
\section*{Acknowledgements}


JSP, JdCP and CMP are funded by projects FPA (MINECO), SGR (Generalitat de Catalunya) and MDM (ICCUB). JdCP also by FPI (MINECO) and CMP by FI (GC). AGV is funded by DFG (Germany).

\bibliographystyle{aasjournal}


\end{document}